\documentclass[]{spie}  

\usepackage[]{graphicx}

\title{Effects of diversification among assets in an agent-based market model}

\author{F. Ghoulmi\'e\supit{a,c}, M. Bartolozzi\supit{a,b}, C.P. Mellen\supit{a}, T. Di Matteo\supit{c}
\skiplinehalf
\supit{a}Research Group, Grinham Managed Funds Pty. Ltd., Sydney NSW 2065, Australia.\\
\supit{b} Special Research Centre for the Subatomic Structure of Matter (CSSM), University of Adelaide, Adelaide SA 5005, Australia.\\
\supit{c}Department of Applied Mathematics, Research School of Physical Sciences and Engineering, The Australian National University, Canberra ACT 0200, Australia.}
\authorinfo{ Fran\c{c}ois Ghoulmi\'e e-mail: Francois.Ghoulmie@gmf.com.au }
 \begin{document}
 \maketitle

\begin{abstract}
We extend to the multi-asset case the framework of a discrete time model of a single asset financial market developed in Ghoulmi\'e et al~\cite{Ghoulmie05}. In particular, we focus on adaptive agents with threshold behavior allocating their resources among two assets. We explore numerically the effect of this diversification as an additional source of complexity in the financial market and we discuss its destabilizing role. We also point out the relevance of these studies for financial decision making.
\end{abstract}


\section{INTRODUCTION}
\label{sect:intro}

In the agent-based approach, financial markets are modelled as
systems of interacting agents and several examples of such
models
have been successful in reproducing the stylized facts that are
common to a wide variety of markets, instruments and
periods~\cite{cont01}. By finding economic explanations for the
statistical signatures of these market fluctuations, agent-based
models can inspire investors and regulators to conceive tools and
policies for improved financial decisions and financial risk
management. In this literature, herd behavior~\cite{cont00,bartolozzi04}, social interaction
and mimicry~\cite{cont00,iori02,stauffer}, heterogeneity~\cite{lebaron01,challet01,licalzi}, investor inertia~\cite{Ghoulmie05,cont00} and switching between ``chartist'' and ``fundamentalist'' behavior~\cite{arthur97,farmer,giardina,kirman,lux} have been invoked as possible mechanisms. The mechanism involved in the single asset
financial market model introduced in Ghoulmi\'e et al~\cite{Ghoulmie05}, adds
clarity on how the stylized facts are generated, and in particular
how the volatility clustering phenomenon is linked to economic
decision making as commented in Cont~\cite{cont05}. In this model,
one type of agent interacts indirectly via the price, and
heterogeneity appears endogenously due to an asynchronous updating
scheme. The structure of the model allows one to trace back the
behavior of the price to agents' decision-making rule based on a
threshold behavior. Threshold response in the behavior of market
participants can be seen either as resulting from trade friction
in order to reduce transactions costs or, more generally, from the
risk aversion of agents which leads them to be inactive if
uncertain about their action.
This agent-based model generically
leads to an absence of autocorrelation in the returns,
mean-reverting stochastic volatility, excess volatility,
volatility clustering, and endogenous bursts of market activity
that is not attributable to external noise. However, the model
does not generate heavy-tails in the distribution of returns or
the long memory effect in the volatility, and this can in itself
motivate a further study by adding complexity to the model in
order to get these facts.
Diversification among multiple assets,
following for example Markovitz's mean-variance portfolio
optimization~\cite{markowitz}, is crucial for investment decisions
and may be the key ingredient to getting all the facts. Moreover,
these stylized facts are so constraining that it is even
surprising to be able to reproduce them in a convincing way with a
single-asset financial market model. The study of the multi-asset
case is also the road to understand the non trivial correlations
between assets~\cite{marsili1,marsili2}, an area of research that
is receiving recently an intense interest with the use of random
matrix theory~\cite{laloux,plerou}, complex
networks~\cite{tiziana1,tiziana2}, and
multi-scaling~\cite{tiziana3,tiziana4,tiziana5,eisler}.
Correlations among returns of different assets are highly
unstable and this is a major challenge for portfolio optimization,
for the pricing of multi-asset derivatives and for co-integration based
trading strategies. Our approach aims also at bringing insights on
the direct impact of trading strategies on speculative markets
dynamics compared to the importance given to microstructure
effects, e.g in the zero-intelligence agents model studied in
Daniels et al~\cite{daniels} and to the importance given to the topology of
interactions between agents in Cont and Bouchaud~\cite{cont00} and Iori~\cite{iori02}. We thus
extend the agent-based model to the multi-asset case and, in order
to obtain transparent pictures, we focus on agents diversifying
their strategies among two assets.

The article is structured as follows. Section \ref{ABM1}
summarizes the single asset model and discusses its main
properties. We describe the multi-asset market model in section
\ref{ABM2}. We discuss the numerical effects of the
diversification in section \ref{op}. Conclusions are drawn in the last
section.

\section{Properties of the single asset market model}
\label{ABM1} The model describes a market where a single asset is
traded by $n$ agents. Trading takes place at discrete time steps
$t$. Provided the parameters of the model are chosen in a certain
range, these periods may be interpreted as ``trading days''. At
each time period, every agent receives public news about the
asset's performance. If the news is judged to be significant the
agent places for a unit of asset a buy or sell order, depending on
whether the news received is pessimistic or optimistic. Prices
then move up or down according to excess demand. The model
produces stochastic heterogeneity and sustains it through the updating of agents' strategies. Let us recall
in a mathematical way the ingredients of the single asset model.
At each time period:\\ \\
$\bullet$ All the agents receive a common signal $\epsilon_t$ generated by a Gaussian distribution with $0$
mean and standard deviation $D$, namely $N(0,D^2)$.\\
$\bullet$ Each agent $i$ compares the signal to its threshold $\theta_i(t)$.\\
$\bullet$ If $|\epsilon_t|>\theta_i(t)$ the agent considers the signal as significant and generates an order $\phi_i(t)$ according to
\begin{equation}
 \phi_i(t)=1_{\epsilon_t > \theta_i}-1_{\epsilon_t < -\theta_i}, \\
\label{decision1}
\end{equation}
where $\phi_i(t)>0$ is a buy order, $\phi_i(t)<0$ is a sell order and $\phi_i(t)=0$ is an order to remain inactive.\\
$\bullet$ The market price $p_t$ is affected by the excess demand and moves according to
\begin{equation}
r_t=\ln\left(\frac{p_{t}}{p_{t-1}}\right) = g\left(\frac{\sum_i
\phi_i(t)}{n}\right), \label{price}
\end{equation}
where $r_t$ are the returns at time $t$ and $g$ is the price impact function.\\
$\bullet$ Each agent updates, with probability $s$, her threshold to $|r_t|$.\\
\\
The evolution of the thresholds distribution is described with the following master equation:
\begin{equation}
f_{t+1}(\theta) =  (1-s)\; f_t(\theta)+s\;  \delta_{|r_t|,\theta},
\label{eq::tha_distribution}
\end{equation}\\
with $|r_t| =|g(\rm sign(\epsilon_t)F_t(|\epsilon_t|))|$, $F_t$ being the cumulative distribution of the thresholds.\\
The solution of Eq.~\ref{eq::tha_distribution} can be derived analytically and reads as\\
\begin{equation}
f_{t}(\theta) =  (1-s)^{t}\;f_0+s\sum_{j=1}^{t} (1-s)^{j-1}\delta_{|r_{t-j}|,\theta}.
\label{evolution}
\end{equation}\\

Moreover, numerical tests confirm the validity of the former
solution.
%
%
 Stationary solutions are the limiting cases:
 without feedback $s=0$, and without heterogeneity $s=1$.
We specify now the range of parameters that leads to realistic
price behaviors. First of all, we want a large number of agents in
order to guarantee heterogeneity in the market. Indeed, when the
number of agents is lowered, the distribution of returns becomes
multi-modal with 3 local maxima, one at zero, one positive maximum
and a negative one. This can be interpreted as a disequilibrium
regime: the market moves either one way or the other. The updating
frequency $s$ should be chosen small, $s<<1$, in order to
guarantee heterogeneity. When the amplitude of the noise is small,
$D<<g(1/n)$, the absolute value of the returns evolves through a
series of periods characterized by ``jumps" whose amplitudes decay
exponentially in time. The sensitivity of the thresholds increases
when the noise level increases and the behavior of the returns is
closer to the Gaussian signal. On the other hand, when the
amplitude of the news is too high, $D>>g(1)$, the returns
distribution has two peaks: the maximum at $g(1)$ and the minimum at $g(-1)$.
 We thus want the following condition in order to get realistic returns dynamics:
\begin{equation}
g(1/n)<<D<<g(1).
\label{condition}
\end{equation}\\
If we consider now the linear price impact function, $g(x)=x/\lambda$, with $\lambda$ the market depth characterizing how much the market moves when filling one unit of asset, the above condition leads to the parameter reduction $D_{\rm eff}=D\lambda$ and $1/n<<D_{\rm eff}<<1$. We then get clusters of volatility of length $1/s$ consistent with the correlation structure suggested by the stationary solution. This slow feedback mechanism generates endogenous heterogeneity, excess volatility, volatility clustering and transforms Gaussian news into semi heavy-tailed price returns. When a majority of agents have a low value for their threshold a large price fluctuation becomes very probable. Because only a small fraction of agents increases its threshold response when a large fluctuation occurs the probability of also getting a large fluctuation at the next time step remains high. In other words, the slow feedback mechanism causes persistence in the fluctuations.

\section{Description of the multi-asset market model}
\label{ABM2}

We now extend the previous model to a two-asset case that can in fact be directly generalized to a higher number of assets. The model describes a market where two assets, with prices denoted by $p_{1,t}$ and $p_{2,t}$ respectively, are traded by $n$ agents
  at discrete time steps $t$.  At each time step, the model is
  updated according to the following steps:\\ \\
$\bullet$ Every agent receives a common signal $\epsilon_t \in N(0,D^{2})$ for both assets.\\
$\bullet$ Each agent $i$ compares the signal to her threshold for
the first, $\theta_{1,i}(t)$, and for the second,
$\theta_{2,i}(t)$, asset and then places orders of respectively $\omega_{1,i}(t)$ and $\omega_{2,i}(t)$ units of assets. The
weights $\omega_{1,i}(t)$ and $\omega_{2,i}(t)$ are defined positive and bounded by the
following constrains:
\begin{equation}
 \omega_{1,i}(t)+\omega_{2,i}(t)=1.\\
\label{constraint_1}
\end{equation}
$\bullet$ If $|\epsilon_t|>\theta_{1,i}(t)$ the agent considers
the signal as significant and generates for the first asset an order $\omega_{1,i}\phi_{1,i}(t)$ with
\begin{equation}
 \phi_{1,i}(t)=1_{\epsilon_t > \theta_{1,i}}-1_{\epsilon_t < -\theta_{1,i}} .\\
\label{eq::phi}
\end{equation}
$\bullet$ If $|\epsilon_t|>\theta_{2,i}(t)$ the agent considers the signal as significant and generates for the second asset an order $\omega_{2,i}\phi_{2,i}(t)$ with
\begin{equation}
 \phi_{2,i}(t)=1_{\epsilon_t > \theta_{2,i}}-1_{\epsilon_t < -\theta_{2,i}} .\\
\label{decision2}
\end{equation}
$\bullet$ The asset prices $p_{1,t}$ and $p_{2,t}$ are affected by the excess demand and move according to
\begin{equation}
r_{1,t}=\ln\left(\frac{p_{1,t}}{p_{1,t-1}}\right) =
g\left(\frac{\sum_i\omega_{1,i}\phi_{1,i}(t)}{n}\right),
\label{price1}
\end{equation}
\begin{equation}
r_{2,t}=\ln\left(\frac{p_{2,t}}{p_{2,t-1}}\right) =
g\left(\frac{\sum_i\omega_{2,i}\phi_{2,i}(t)}{n}\right),
\label{price2}
\end{equation}
where $r_{1,t}$ and $r_{2,t}$ are the returns and $g$ is the price impact function.\\
$\bullet$ Each agent updates, with probability $s$, its threshold $\theta_{1,i}(t)$ to $|r_{1,t}|$ and $\theta_{2,i}(t)$ to $|r_{2,t}|$.\\
\\
\\
$\bullet$ Each agent updates, with probability $s$, the weights according to
\begin{equation}
\omega_{1,i}(t)=\frac{1}{1+e^{\beta.V_t}}, \label{weight1}
\end{equation}
\begin{equation}
\omega_{2,i}(t)=1-\omega_{1,i}(t), \label{weight2}
\end{equation}
where the difference between assets absolute returns $V_t=|r_{2,t}|-|r_{1,t}|$ is the fitness
function that determines the adaptive asset allocation and $\beta$ is the parameter
that controls the intensity of choice between the two assets.\\
The asset allocation follows a discrete choice model that is
popular among economists under the label {\em logit} choice
model~\cite{mcfadden}. It has been used in various models of
social choices~\cite{durlauf1} and it is an example of the intense
and fertile exchange between physics and social sciences as
explained and reviewed in Nadal et al~\cite{nadal}. The parameter $\beta$
determines the rationality of agent's choice. For $\beta=0$,
agents keep the same equal weights on both assets independently of
the fitness function, and the behavior of the returns is
equivalent to the single asset market studied in the first
section. For an infinite value of $\beta$, agents are fully
rational and chose radically the most profitable asset, the asset
with the highest absolute return.

\section{Numerical effects of the diversification}
\label{op}

\subsection{Nonstationarity of assets returns}
\label{ns}

\begin{figure}
\vspace{2cm}
\begin{center}
\begin{tabular}{c}
\includegraphics[height=8cm,width=14cm]{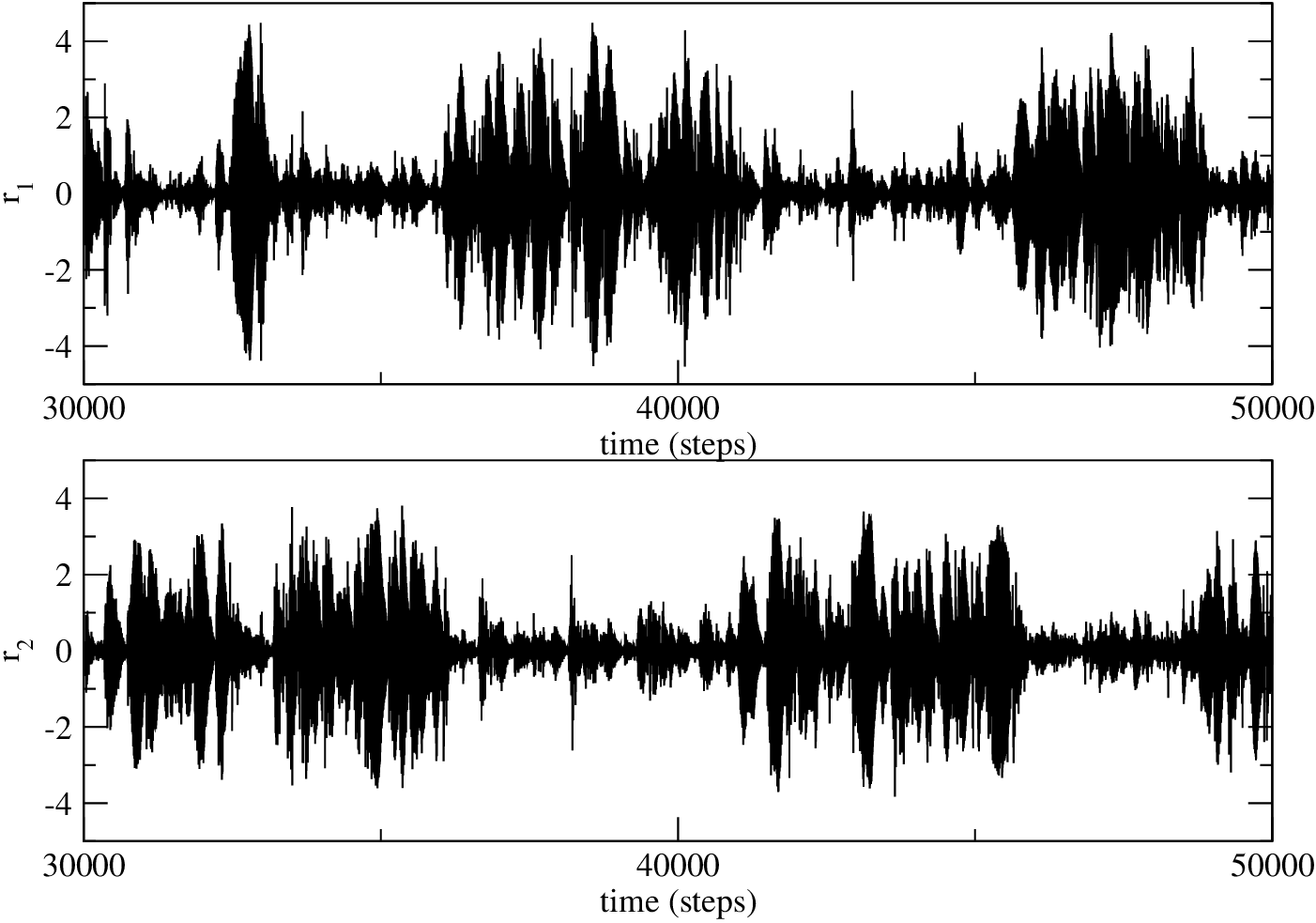}
\end{tabular}
\end{center}
\caption{Time series plots of asset1 and asset2 returns, $r_1$
(top) and $r_2$ (bottom), generated numerically by the agent-based
market model for $n=1500$, $D=0.001$, $\lambda=2$, $s=0.015$ and
$\beta=1000$. The time series exhibit regime-switching type of
behaviors between a high and low volatility periods.}
\label{fig::g1000returns}
\end{figure}

\begin{figure}
\vspace{2cm}
\begin{center}
\begin{tabular}{c}
\includegraphics[height=8cm,width=14cm]{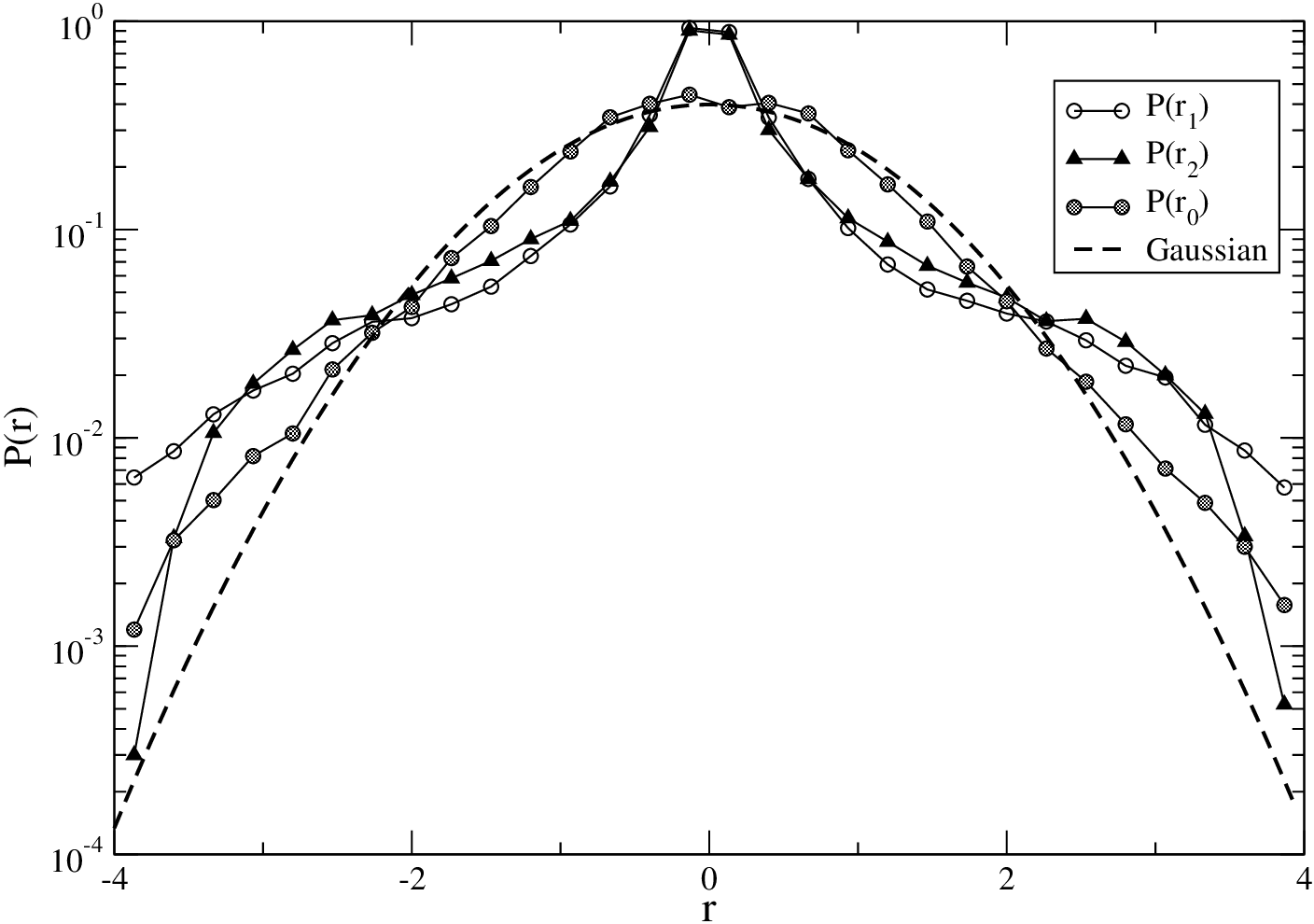}
\end{tabular}
\end{center}
\caption{Empirical densities plots $P(r_1)$ and $P(r_2)$ for
asset1 and asset2 time series returns, generated numerically by
the agent-based market model for $n=1500$, $D=0.001$, $\lambda=2$,
$s=0.015$ and $\beta=1000$. These time series exhibit fatter tails
than the Gaussian distribution. The Gaussian distribution case and
the empirical density $P(r)$ for the returns generated by the
single-asset model for $n=1500$, $D=0.001$, $\lambda=10$ and
$s=0.015$ are also plotted for comparison. }
\label{fig::pdf_g1000returns}
\end{figure}

In this section we explore numerically the effects of the
diversification on the complex behavior of asset returns. In
particular, by setting the other parameters
in ranges similar to those required to get realistic dynamics in the single asset case, we study the role of the parameter $\beta$ that
characterizes the asset allocation. The price impact function has
been chosen to be linear, $g(x)=x/\lambda$, with the same market
depth $\lambda$ for both assets. The behavior of the returns
generated by the present model is significantly different from the
single asset case. In particular, the returns, reported in
Fig.~\ref{fig::g1000returns} for $n=1500$, $s=0.015$, $D=0.001$,
$\lambda=2$ and $\beta=1000$, produce fluctuations that are larger than those of the
the single asset model. This behavior is emphasized
by the fatter tails observed in the probability distribution
function and reported in Fig.~\ref{fig::pdf_g1000returns} where a
Gaussian distribution is also plotted for visual comparison.

\begin{figure}
\vspace{1.7cm}
\begin{center}
\begin{tabular}{c}
\includegraphics[height=8cm,width=14cm]{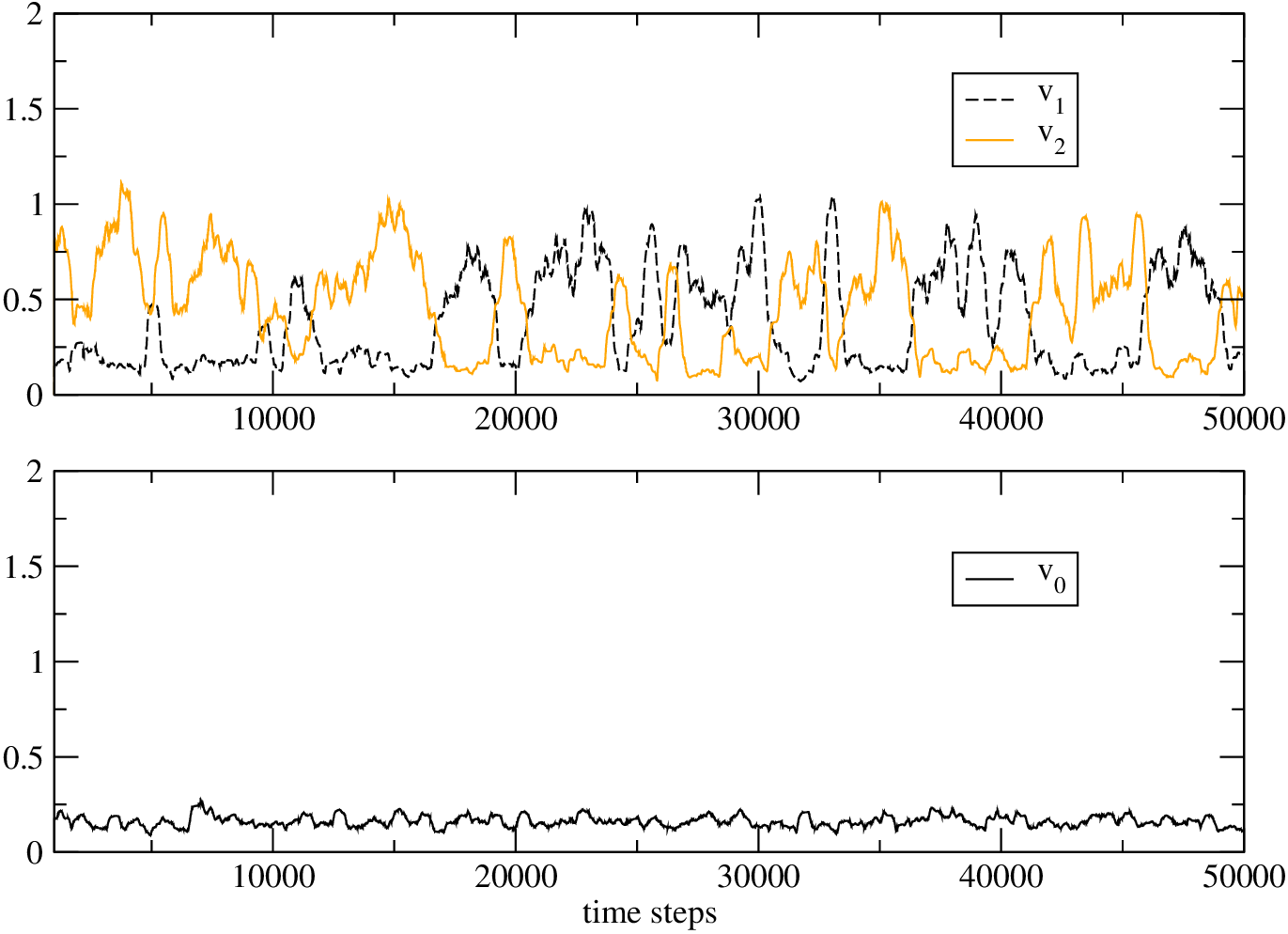}
\end{tabular}
\end{center}
\caption{Top: Annualized volatilities $v_{1}$ and $v_{2}$
for the time series returns of the two assets generated
numerically by the agent-based market model with $n=1500$,
$D=0.001$, $\lambda=2$, $s=0.015$ and $\beta=1000$. Bottom:
Annualized volatility $v_0$ for the returns generated by the
single-asset model for $n=1500$, $D=0.001$, $\lambda=10$ and
$s=0.015$. The volatilities are calculated as the standard
deviation of the returns on a moving window of 500 time steps
multiplied by the annualization factor $\sqrt{250}$.}
\label{fig::annualized_volatility}
\end{figure}

Fig.~\ref{fig::annualized_volatility} illustrates that additional differences from the single asset model are observed in the behavior of the annualized volatilities. It is seen that the volatilities of the two-asset model switch intermittently, in a sort of mean reverting
behavior, from high to low values in an anticorrelated fashion. These sudden changes in the volatility are observed in financial data and in order to capture this feature, extensions of GARCH models with regime-switching type of behavior have been proposed such as in Bauwens et al~\cite{bauwens}. Compared to the single asset model, this new result brings us closer to the empirical facts observed in financial data. Moreover, these higher fluctuations occur
without structural changes in the market or as a result of changes
in  the fundamentals and are simply related to the diversification process: the agents having very similar trading
strategies allocate intermittently more weight on one of the asset
and, therefore, leading to higher fluctuations in the traded volume.

We also perform an analysis of summary statistics for 5 consecutive periods of
10000 time steps for the time series generated by the multi-asset model
and the single-asset model. These summary statistics are reported
in Tabs.~\ref{sumstat1}, \ref{sumstat2} and \ref{sumstat3} respectively. In the
two-asset case we observe fluctuations of the kurtosis between 3
and 12, indicating nonstationarity of asset returns. In the single
asset case, instead, this parameter is almost stationary. This result implies
that the multi asset diversification of the proposed model promotes non-stationarity, an effective feature that often causes problems to policymakers.
We must remark on the
destabilizing role of this diversification in comparison with the
benchmark single asset market case. This is an interesting case
where diversification, which is usually seen as a way to reduce portfolio risk, increases
market instability.

\begin{table}[htbp]\centering
    \centering
        \begin{tabular}{l c c c c c c }\hline\hline
\multicolumn{1}{c}{\textbf{Period}} & \textbf{mean($10^{-4}$)} &
\textbf{std($10^{-2}$)} & \textbf{skew($10^{-2}$)} &
\textbf{kurtosis} & \textbf{max($10^{-2}$)} & \textbf{min($10^{-2}$)}\\
\hline
1 & 1.1 & 1.3 & 24 & 11.1  & 10.19 & -9.64\\
2 & 7.3 & 2.3 & 7.5 & 7.6 & 12 & -12.53\\
3 & 0.6 & 4 & -1.9 & 4 & 13.44 & -13.38\\
4 & 3.9 & 3.1 & 1.9 & 6.5 & 12.87 & -12.94\\
5 & -2.4 & 2.8 & 0.5 & 5.7 & 12.29 & -12.89\\
        \end{tabular}
        \caption{Summary statistics for the returns of the first asset in the two-asset model.
        The estimation is obtained over 10000 observations.}
        \label{sumstat1}.
\end{table}

\begin{table}[htbp]\centering
    \centering
        \begin{tabular}{l c c c c c c }\hline\hline
\multicolumn{1}{c}{\textbf{Period}} & \textbf{mean($10^{-4}$)} &
\textbf{std($10^{-2}$)} & \textbf{skew($10^{-2}$)} &
\textbf{kurtosis} & \textbf{max($10^{-2}$)} & \textbf{min($10^{-2}$)}\\
\hline
1 & 2.2 & 4.3 & 8.3 & 3.5  & 13.37 & -13.63\\
2 & 6.8 & 3.6 & 4 & 4.3 & 12.85 & -12.89\\
3 & 0.6 & 1.8 & -7 & 11.6 & 12.71 & -12.33\\
4 & 3.8 & 3.1 & 2.5 & 5.7 & 12.61 & -11.89\\
5 & -0.1 & 3.2 & 3 & 5.6 & 12.1 & 12.63\\
        \end{tabular}
        \caption{Summary statistics for the returns of the second asset in the two-asset
        model. The estimation is obtained over 10000 observations.}
        \label{sumstat2}
\end{table}

\begin{table}[htbp]\centering
    \centering
        \begin{tabular}{l c c c c c c }\hline\hline
\multicolumn{1}{c}{\textbf{Period}} & \textbf{mean($10^{-4}$)} &
\textbf{std($10^{-2}$)} & \textbf{skew($10^{-2}$)} &
\textbf{kurtosis} & \textbf{max($10^{-2}$)} & \textbf{min($10^{-2}$)}\\
\hline
1  & 0.05 & 1.05 & 4.3 & 4.3  & 6.7 & -5.3\\
2  & 0.5 & 1.04 & 4.2 & 4.2 & 5.13 & -5.4\\
3  & 0.03 & 1.05 & 3.6 & 3.6 & 4.4 & -5.2\\
4  & 2.7 & 1.05 & 3 & 4.7 & 5.3 & -5.6\\
5  & 0.4 & 1.01 & -5 & 4 & 5.3 & -6.7\\
        \end{tabular}
         \caption{Summary statistics for the returns in the single asset model. The estimation is obtained over 10000
         observations.}
         \label{sumstat3}
\end{table}

\subsection{Nonlinear diagnostics}
\label{sec::nl}

\begin{figure}
\vspace{2cm}
\begin{center}
\begin{tabular}{c}
\includegraphics[height=8cm,width=14cm]{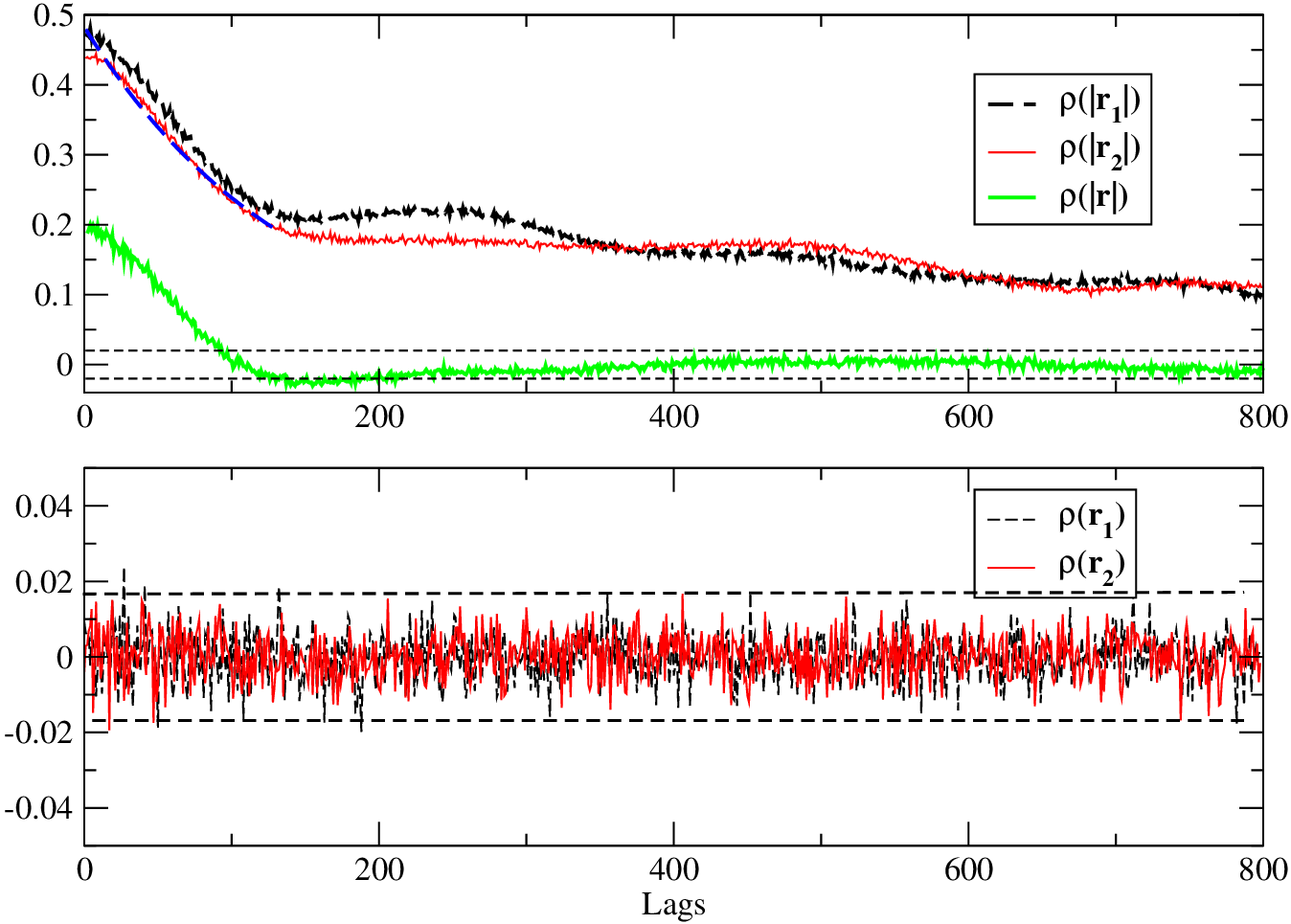}
\end{tabular}
\end{center}
\caption{Top: Autocorrelation functions $\rho(|r_1|)$ and
$\rho(|r_2|)$ of $|r_1|$ and $|r_2|$  generated numerically
by the agent-based market model for $n=1500$, $D=0.001$,
$\lambda=2$, $s=0.015$ and $\beta=1000$. The volatility clustering phenomenon is reproduced: these functions remain significantly positive over a long period with an initial phase similar to the
single asset case. An exponential fit (dashed line) of the initial decay has been plotted. We also plotted for comparison the
autocorrelation function $\rho(|r|)$ for the absolute values of the
returns generated by the single-asset model with $n=1500$,
$D=0.001$, $\lambda=10$ and $s=0.015$. In the two-asset case, we observe at large lags a slower decay and the autocorrelations remain significantly positive. The persistence in the amplitude of the returns in the two-asset case is definitely closer to that observed in financial data than is the persistence generated in the one-asset case. Bottom: Corresponding
autocorrelation functions $\rho(r_1)$ and $\rho(r_2)$ of $r_1$ and
$r_2$. The multi-asset model generates an absence of
autocorrelation in the returns.} \label{fig::ACF}
\end{figure}

In the present section we estimate the autocorrelation function
for the returns of the two assets as well as their absolute value.
The results, reported in Fig.~\ref{fig::ACF}, show an absence of
correlation in the returns time series (bottom). For the absolute
value (top), the nonlinear diagnostics point out two
different phases: the first, for small lags, is related to the presence
of volatility clustering, a feature noticed also in the single asset
model, the second persistent phase has not previously been observed. The
first phase, analogously to that seen in the one asset model, corresponds  to
the clusters of volatility of length $1/s$ and can be described by
an exponential decay with characteristic time scale of 70 periods.
At large lags we instead observe a slower decay while the autocorrelations remain significantly positive over a long
period.  We thus conclude that the fluctuations persist in the
multi-asset case over a longer period than in the single asset
case, Fig.~\ref{fig::ACF}. This behavior is definitely closer to
that observed in financial time series than is the behavior exhibited by the
single asset model.


\subsection{Robustness and implications of the results}
\label{rob}

\begin{figure}
\vspace{2.5cm}
\begin{center}
\begin{tabular}{c}
\includegraphics[height=8cm,width=14cm]{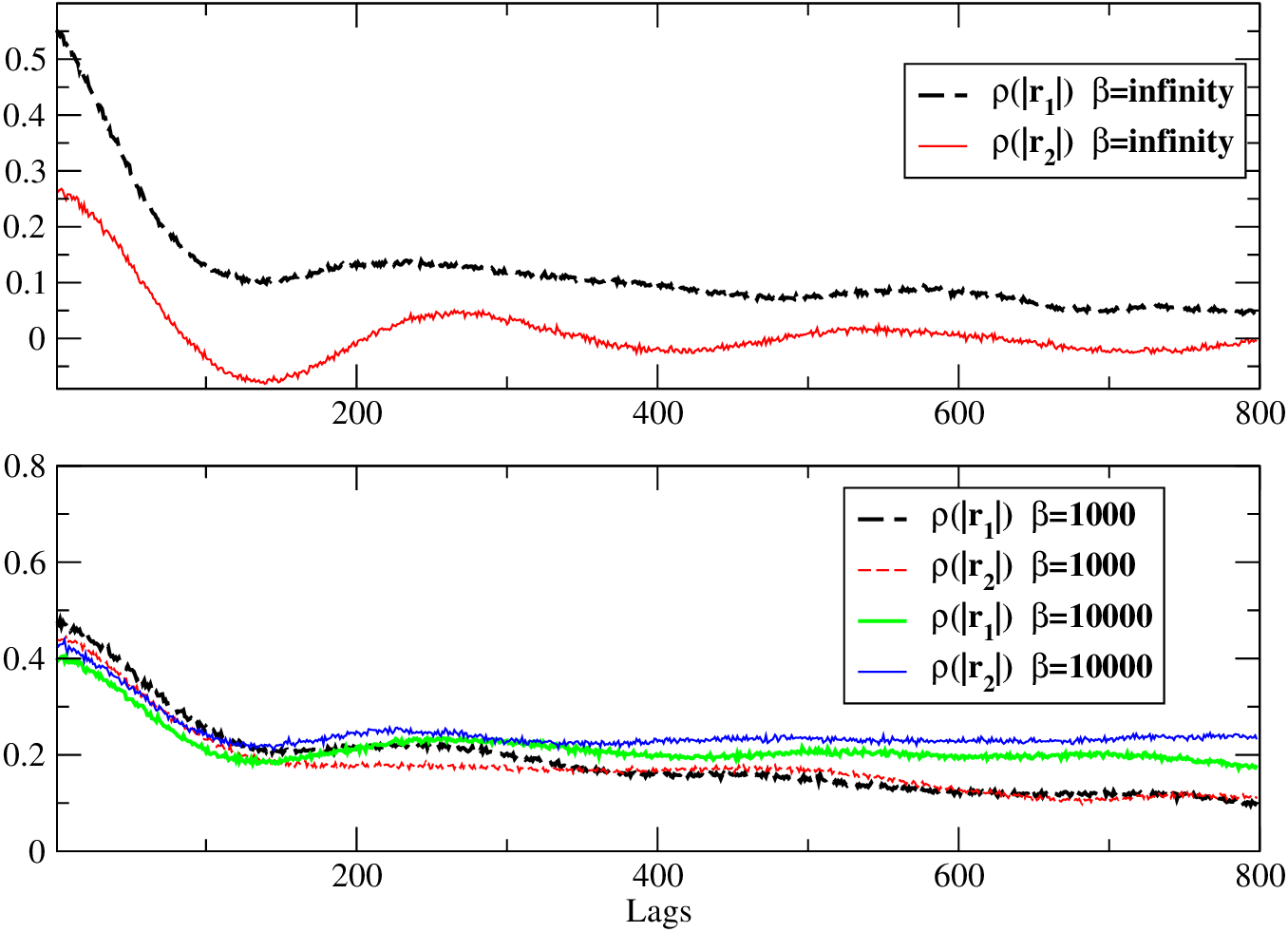}
\end{tabular}
\end{center}
\caption{Top: Autocorrelation functions $\rho(|r_1|)$ and
$\rho(|r_2|)$ of $|r_1|$ and $|r_2|$, absolute values of asset1 and
asset2 time series returns generated numerically by the
agent-based market model for $n=1500$, $D=0.001$, $\lambda=2$,
$s=0.015$ and an infinite value of $\beta$. This limit case, which
is the radical choice case, exhibit two distinct behaviors for the
volatilities of the two assets returns. Bottom: Autocorrelation
functions $\rho(|r_1|)$ and $\rho(|r_2|)$ of $|r_1|$ and $|r_2|$,
absolute values of the two assets time series returns generated
numerically by the agent-based market model for $n=1500$,
$D=0.001$, $\lambda=2$, $s=0.015$, $\beta=1000$ and $\beta=10000$.
When varying the intensity of choice, the model is robust in
generating higher persistence in the fluctuations compared to the
single asset case.} \label{fig::ACF_2}
\end{figure}

A parametric analysis of the two-asset model confirms the
robustness of the results presented in the previous two sections:
this is a desirable property for practitioners when building market
models. Varying the updating frequency, $s$, affects the initial
exponential decay of the autocorrelation function of the
volatilities. The characteristic time of this decay is, indeed,
roughly proportional to $1/s$. When increasing significantly the
number of agents, $n$, we have to decrease the parameter $D_{\rm eff}$
in order to get an effective impact of the asset allocation on
market moves and reproduce the results reported in the previous
section. When increasing the intensity of choice, $\beta$, the
diversification is more effective and the results are more
pronounced: for $\beta=10000$, the fluctuations in the kurtosis of
the returns is higher as are the fluctuations in $V_t$. The
autocorrelation function of the instantaneous volatility, instead,
remains unchanged against variations of $\beta$ for both assets, see
Fig.~\ref{fig::ACF_2}. The limit case with infinite $\beta$
exhibits two distinct behaviors for the volatilities of the two
assets as a result of the radical choice: returns similar to the
single asset case for the most often chosen asset and returns with
lower amplitude but with higher variations and persistence in the
volatility for the other. The autocorrelation function for this
particular case is reported in Fig.~\ref{fig::ACF_2} where it is
evident that the volatility behaves similarly to the single asset
case which is plotted alongside.

By using a comprehensive and dynamic approach in the asset allocation
decision this agent-based model also offers a plausible
mechanism for generating correlations among returns of different
assets. The model is successful in reproducing an observed
phenomenon in speculative markets where traders crowd or flock
from one volatile security to another. Moreover, we computed  the
correlation coefficient between assets returns for 5 consecutive
periods of 10000 time steps generated by the multi-asset model.
This correlation parameter fluctuates between 0.3 and 0.7 and we
conclude that the correlations are unstable. This instability
emerges endogenously as a result of trading strategies and not
from the input signal that can be interpreted as the fundamentals
of these assets.  We note the relationship between volatilities
and correlations.  The correlation parameter is indeed positively
correlated to the difference between volatilities which determines
the asset allocation choice and its impact on market movements. The
agent-based approach can thus lead to a better appreciation of the
relationships among many asset classes, and this should improve
the financial decision-making process.

\section{Discussions and Conclusions}

In the present work we have extended to the multi-asset case a
parsimonious single asset agent-based market model capable of reproducing the main empirical
stylized facts observed in the returns of financial assets.\\
The model is based on three main ingredients:\\ \\
$\bullet$ Threshold behavior of agents.\\
$\bullet$ Heterogeneity of agent strategies, generated endogenously through a feedback effect of recent price.\\
$\bullet$ Diversification among assets following a discrete choice model.\\

Compared to the single asset case, numerical simulations of the model produce higher
 variations of assets returns volatilities and an autocorrelation function of absolute returns
 that remains positive over a longer period. Moreover, the model mimics a regime switching type
 in assets dynamics along with an endogenous instability
 in assets correlations as the result of trading strategies and without any structural change. These results question some conclusions
 previously drawn from simulations of agent-based models regarding the origins of stylized properties
 of asset returns, e.g the long memory effect, and also question the importance given to microstructure effects
 and assets fundamentals in previous studies. Taking into account diversification among several assets,
 a key feature of financial decision making, when designing agent-based market models may be critical in
generating all the stylized facts. The model is also successful at
reproducing the crowd dynamics between volatile assets observed on
trading floors. Traders do indeed flock from one volatile security
to another one, a phenomenon that has not been extensively
explored in agent-based modelling. These cross-market dynamics are
not only the road to fully understand the instability  of assets
correlations, but may also explain the persistence in the returns
volatilities and the  heavy-tailed distribution of the returns. We
will thus explore this path and other cross-market dynamics in
further studies.  For practitioners, the potential direct benefits
from understanding  these fluctuations are improved estimates of
the quantities used for trading strategies,  portfolio
optimization, derivatives pricing and risk management. By shedding
light on the origins of market movements, the agent-based approach
can inspire new strategies and policies for traders and
regulators.  It is also to sophisticated investors a first class
guide for understanding the ecology of financial markets.\\

\acknowledgments     
The authors would like to thank Richard Grinham for careful reading. F.G. and T.D.M would also like to thank T. Aste, F. Clementi, R. Cont and J.-P. Nadal for fruitful discussions and advice. T.D.M. acknowledges partial support from COST P10 "Physics of Risk" project and ARC Discovery Projects: DP03440044 (2003) and DP0558183 (2005).\\

\end{document}